\documentclass[amssymb,amsmath,floatfix,pre, twocolumn,superscriptaddress,nolongbibliography]{revtex4-1}
\usepackage{graphicx}
\usepackage{dcolumn}
\usepackage{bm}
\usepackage{hyperref,color}
\usepackage{xcolor}

\newcommand{\tcal}{\mathcal{T}}
\graphicspath{{./figs/}}

\begin{document}

\title{Relaxation-speed crossover in anharmonic potentials}

\author{Jan Meibohm}
\affiliation{Department of Physics and Materials Science, University of Luxembourg, L-1511 Luxembourg, Luxembourg}
\author{Danilo Forastiere}
\affiliation{Department of Physics and Materials Science, University of Luxembourg, L-1511 Luxembourg, Luxembourg}
\author{Tunrayo Adeleke-Larodo}
\affiliation{Department of Physics and Materials Science, University of Luxembourg, L-1511 Luxembourg, Luxembourg}
\author{Karel Proesmans}
\affiliation{Department of Physics and Materials Science, University of Luxembourg, L-1511 Luxembourg, Luxembourg}
\affiliation{Hasselt University, B-3590 Diepenbeek, Belgium}

\begin{abstract}
In a recent Letter [A.~Lapolla and A.~Godec, Phys. Rev. Lett. {\bf 125}, 110602 (2020)], thermal relaxation was observed to occur faster from cold to hot (heating) than from hot to cold (cooling). Here we show that overdamped diffusion in anharmonic potentials generically exhibits both faster heating and faster cooling, depending on the initial temperatures and on the potential's degree of anharmonicity. We draw a relaxation-speed phase diagram that localises the different behaviours in parameter space. In addition to faster-heating and faster-cooling regions, we identify a crossover region in the phase diagram, where heating is initially slower but asymptotically faster than cooling. The structure of the phase diagram is robust against the inclusion of a confining, harmonic term in the potential as well as moderate changes of the measure used to define initially equidistant temperatures.
\end{abstract}
\maketitle
Many thermal relaxation processes in nature and industry occur out of equilibrium, and thus outside of the realm of the quasistatic approximation. As a consequence, nonequilibrium thermal relaxation gives rise to anomalous effects, such as ergodicity breaking~\cite{bray2002theory} or the Mpemba effect~\cite{mpemba1969cool}. The latter describes the surprising observation that some systems cool down faster, when relaxing from a higher initial temperature. A better understanding of such anomalous relaxation effects in out-of-equilibrium systems is important, because it may allow us to use these phenomena to our advantage, for instance, for increasing the rate of heating and cooling.

Although a complete understanding of anomalous relaxation in macroscopic systems appears elusive at present, much progress has been made recently in reproducing anomalous relaxation phenomena on mesoscopic scales. This has led to several important results such as new theoretical~\cite{lu2017nonequilibrium,klich2019mpemba,walker2021anomalous,chetrite2021metastable} and experimental~\cite{kumar2020exponentially,kumar2021anomalous} insights into the Mpemba effect, strategies to increase the rate at which systems can be cooled~\cite{gal2020precooling,prados2021optimizing,carollo2021exponentially}, and an information-theoretic bound on the speed of relaxation to equilibrium~\cite{shiraishi2019information}.

Within a setup closely related to, yet slightly different from, the Mpemba effect, a recent study~\cite{lapolla2020faster} reported an asymmetry in the rate at which systems heat up and cool down. According to this study, and subsequent works by other authors, heating occurs faster than cooling for diffusive systems with harmonic potentials~\cite{lapolla2020faster} and for discrete-state two-level systems~\cite{manikandan2021faster,van2021toward}. On the other hand, it was shown that this relaxation asymmetry is non-generic for diffusion in potentials with multiple minima~\cite{lapolla2020faster} or in discrete-state systems with more than two states~\cite{manikandan2021faster,van2021toward}. However, it appears to be widely believed that the described effect is a general property of overdamped, diffusive systems with stable single-well potentials~\cite{lapolla2020faster,manikandan2021faster,van2021toward}.

In this Letter, we study the relaxation asymmetry for overdamped diffusion in anharmonic potentials. Opposing common belief, we show that these systems exhibit both behaviours, faster heating and faster cooling, even for stable single-well potentials. Based on these results, we draw a phase diagram locating the different regions of ``faster heating'' and ``faster cooling'' in parameter space. These two regions are separated by a crossover region where cooling occurs faster at first, but heating overtakes at a finite time. Our results suggest that the relative speed of thermal relaxation to equilibrium can be substantially increased by varying the anharmonicity of the potential. This should be testable in experiments and has potential applications in the optimisation of cooling strategies for small-scale systems~\cite{gal2020precooling}.

To specify the problem, consider two equilibrium systems, otherwise identical, but at different temperatures $T_c<T_h$. We call the system at temperature $T_c$ cold and that at temperature $T_h$ hot. At time $t=0$, both systems experience an instantaneous temperature quench to the same final temperature $T_f$, where $T_c<T_f<T_h$. The relaxation of the two systems toward equilibrium is monitored by their nonequilibrium free-energy difference~\cite{lapolla2020faster},
\begin{align}\label{eq:neqf}
    \mathcal{F}_i(t)=&\left\langle \Delta E_i(t)\right\rangle- T_f\left\langle \Delta S_i(t)\right\rangle\,,\nonumber\\
    =& k_\text{B} T_f\int_{-\infty}^\infty \!\!\text{d}x\, p_i(x,t)\ln\left[\frac{p_i(x,t)}{p_f(x)}\right]\,,
\end{align}
with respect to the equilibrium distribution $p_f$ at final temperature $T_f$. Here, $\left\langle \Delta E_i(t)\right\rangle$ and $\left\langle \Delta S_i(t)\right\rangle$ are the average differences in the energy and entropy of the (cold or hot) system at time $t$ and its equilibrium state at temperature $T_f$; $k_\text{B}$ denotes the Boltzmann constant. The index $i$ in Eq.~\eqref{eq:neqf} takes the values $c$ and $h$, and $p_c$ and $p_h$ denote the probability densities of the initially cold and hot system, respectively.

In order to quantitatively compare the distances $\mathcal{F}_i(t)$ from equilibrium, the temperatures $T_c$ and $T_h$ at $t=0$ are chosen so that $\mathcal{F}_c(0) = \mathcal{F}_h(0)$~\cite{lapolla2020faster}. We call such a temperature quench ``$\mathcal{F}$-equidistant,'' i.e., at equal distance with respect to the temperature measure \eqref{eq:neqf}. A comparison between this setup and the Markovian Mpemba effect~\cite{lu2017nonequilibrium,klich2019mpemba} is made in Sec.~I of the Supplemental Material (SM)~\cite{SM}.

The specific measure \eqref{eq:neqf} is used for two reasons. First, $\mathcal{F}_i$ is a thermodynamic quantity for systems at equilibrium and hence for $t<0$ and in the limit $t\to\infty$. Second, it remains well defined out of equilibrium and thus for all finite times $t$.

In the long-time limit, both the cold and the hot system relax to equilibrium so that $\mathcal{F}_c(t)$ and $\mathcal{F}_h(t)$ tend to zero asymptotically. The relative distance from equilibrium of the two systems is conveniently measured by the logarithmic ratio
\begin{align}\label{eq:rlog}
	\mathcal{R}(t) \equiv \ln\left[	\frac{ \mathcal{F}_h(t)}{\mathcal{F}_c(t)}	\right]\,.
\end{align}
For overdamped diffusion in a harmonic potential, one can prove that $\mathcal{R}(t)>0$, i.e., $\mathcal{F}_h(t)>\mathcal{F}_c(t)$ during the relaxation~\cite{lapolla2020faster}, i.e., heating occurs faster than cooling; $\mathcal{R}(t) < 0$ corresponds to the opposite case, that of faster cooling. Note also that $\mathcal{R}(0)=0$ by definition of $\mathcal{F}$ equidistance, $\mathcal{F}_c(0) = \mathcal{F}_h(0)$. Hence, the momentary, relative distance from equilibrium is determined by the sign of $\mathcal{R}(t)$.

We study the evolution of $\mathcal{R}(t)$ for overdamped diffusion in an anharmonic potential $V(x)$. For simplicity, we analyse the case of one spatial dimension and assume $V(x)$ to be of the form $V(x) = \lambda x^2 + k |x|^{\alpha}$, where we consider parameter values $\lambda$, $k$ and $\alpha$ for which $V$ is confining, $V(x) \to \infty$ as $x\to\pm\infty$. 
We move to a dimensionless formulation by defining a timescale $\tau$ and a length scale $\ell$ as
\begin{equation}
    \tau = \frac{1}{\mu k_\text{B} T_f} \left(\frac{k}{k_\text{B} T_f}\right)^{-2/\alpha}\,,\quad \ell = \left( \frac{k}{k_\text{B} T_f}\right)^{-1/\alpha}\,.
\end{equation}
Here, $\mu$ is the mobility. In the dimensionless coordinates, times are measured in units of $\tau$, lengths in units of $\ell$, and energies in units of $k_\text{B} T_f$. In particular, the transformation $t\to \tilde t = t/\tau, x\to \tilde x = x/\ell$, $V(x)\to \tilde V=V/(k_BT_f)$, to dimensionless coordinates yields, after dropping the tildes, the potential
\begin{equation}\label{eq:pot}
    V(x) = \sigma x^2 + |x|^\alpha\,,
\end{equation}
with the dimensionless parameter $\sigma=\lambda k^{-2/\alpha}(k_\text{B}T_f)^{2/\alpha-1}$. The parameter $\sigma$ quantifies the importance of the harmonic term $x^2$ compared to the anharmonic term $|x|^\alpha$. We focus here on either monomial potentials with $\sigma=0$, or on the case where $\sigma$ is small. Small $\sigma$ occurs whenever (1) $\lambda$ is small, i.e, the harmonic coupling is weak, or (2) $k$ is large, corresponding to strong anharmonic coupling. In addition, one has the cases (3) $0<\alpha<2$ and small $T_f$, where the behaviour is dominated by the (anharmonic) shape of the potential close to the origin, and (4) $\alpha>2$ and large $T_f$, i.e., the dynamics takes place in the anharmonic wings of the potential $V(x)$.

The  Fokker-Planck equation~\cite{risken1996fokker} that determines the evolution of the probability density during the relaxation reads, in the new coordinates, $\partial_t p_i(x,t) = \mathcal{L} p_i(x,t)$ with
\begin{align}\label{eq:fp}
    \mathcal{L} = \partial_x\left[V'(x) + \partial_x \right]\,,
\end{align}
and initial conditions,
\begin{align}\label{eq:init}
    p_i(x,0) =  \frac{\exp[-V(x)/{\tcal_i}]}{Z_{\tcal_i}}\,.
\end{align}
Here, we introduced the dimensionless temperature ratios $\tcal_i$ that are either $\tcal_h \equiv T_h/T_f$ or $\tcal_c \equiv T_c/T_f$, depending on whether the initial temperature before the quench is $T_h$ or $T_c$. Note that for the final-temperature ratio $\tcal_f \equiv T_f/T_f=1$. The constants $Z_{\tcal_i}$ in Eq.~\eqref{eq:init} are obtained from normalising the probability density. 

In the limit $t\to\infty$, the densities $p_i(x,t)$ relax to the equilibrium distribution, $p_f(x) = \exp[-V(x)]/Z_1$. Hence, after the $\mathcal{F}$-equidistant temperature quench at $t=0$, the evolution of the relative distance from equilibrium, measured by $\mathcal{R}(t)$ [Eq.~\eqref{eq:neqf}], is a function of the parameters $\sigma$, and $\alpha$ of the potential $V(x)$ [Eq.~\eqref{eq:pot}] and of the temperature ratios $\tcal_i$ that enter in the initial conditions \eqref{eq:init}. 

\begin{figure}
    \includegraphics[width=\linewidth]{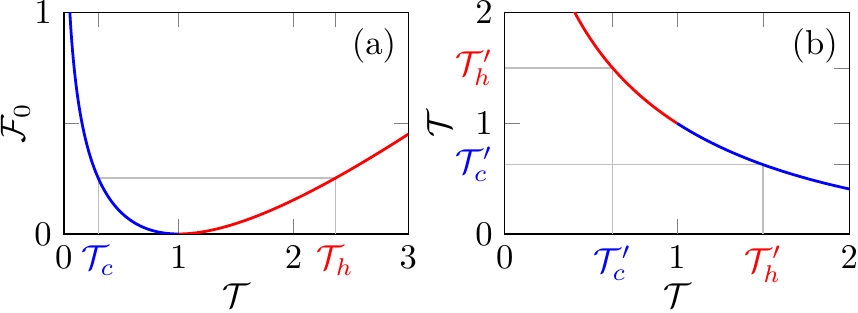}
    \caption{(a) Free-energy difference $\mathcal{F}_0$ at time $t=0$ for hot (red line) and cold (blue line) temperatures, $\tcal_h$ and $\tcal_c$, respectively. The $\mathcal{F}$-equidistance relation \eqref{eq:equidistance} is represented by the grey lines. (b) $\tcal_h(\tcal_c)$ (red line) and $\tcal_c(\tcal_h)$ (blue line) for $\sigma=0$, Eqs.~\eqref{eq:tauhc}. The grey lines and coloured labels indicate a temperature pair $(\tcal'_c,\tcal'_h)$, related by $\mathcal{F}$ equidistance.}
    \label{fig:tauhc}
\end{figure}

Prior to the temperature quench, the hot and cold systems are prepared at $\mathcal{F}$ equidistance so that their free-energy differences match. This condition implicitly relates the hot and cold temperature ratios, so that we can write $\tcal_c(\tcal_h)$, with
\begin{equation}\label{eq:equidistance}
	\mathcal{F}_h(0) = \mathcal{F}_c(0)\big|_{\tcal_c(\tcal_h)}\equiv     \mathcal{F}_0\,.
\end{equation}
Because $\mathcal{F}$ has a single minimum at equilibrium where $\tcal=\tcal_f=1$ and $\mathcal{F}=0$, there is always exactly one solution to Eq.~\eqref{eq:equidistance} for which $\tcal_c(\tcal_h)<1<\tcal_h$. Figure~\ref{fig:tauhc}(a) shows schematically how the free-energy difference relates the different temperatures. 

At $t=0$, the formula for the dimensionless free energy difference $\mathcal{F}_0$ at equidistance [Eq.~\eqref{eq:neqf} in units of $k_\text{B} T_f$] can be conveniently written as
\begin{align}\label{eq:f0}
	\mathcal{F}_0   =& \left[1 + (1-\tcal)\partial_{\tcal^{-1}}\right]\ln\left(\frac{Z_{\tcal}}{Z_1}\right)\,,
\end{align}
where $\tcal=\tcal_h$ when $\tcal>1$ and $\tcal=\tcal_c$ when $\tcal<1$. Hence, in order to obtain the required $\mathcal{F}$-equidistant temperatures, we need to solve and invert Eq.~\eqref{eq:f0}. This can be done analytically for $\sigma=0$, where we find 
\begin{align}\label{eq:f0sol}
	\mathcal{F}_0 = \frac1\alpha\left[\tcal -(1 + \ln\tcal)	\right]\,,
\end{align}
and by taking the inverse
\begin{equation}\label{eq:tauhc}
    \tcal_h = -W_{-1}\left(-\tcal_c e^{-\tcal_c} \right)\,,\quad
    \tcal_c = -W_0\left(-\tcal_he^{-\tcal_h} \right)\,.
\end{equation}

Here, $W_n(x)$, $n=-1,0$ denotes Lambert (or product-log) function \cite{dlmf}. Figure~\ref{fig:tauhc}(b) shows $\tcal_h(\tcal_c)$ (red line) and $\tcal_c(\tcal_h)$ (blue line) from Eqs.~\eqref{eq:tauhc}. For $\sigma\neq0$ the implicit condition \eqref{eq:equidistance} must to be inverted numerically but the curves remain almost unchanged (not shown).

After preparing the hot and cold systems at $\mathcal{F}$-equidistant temperatures, both systems are put in contact with the same heat bath with $\tcal_f = 1$. At finite time $t>0$, the probability densities $p_i(x,t)$ that enter $\mathcal{F}_i(t)$ and thus $\mathcal{R}(t)$ are obtained from the Fokker-Planck equation by
\begin{align}\label{eq:evolution}
	p_{i}(x,t) = e^{\mathcal{L}t}p_i(x,0)\,.
\end{align}
In other words, in order to compute $\mathcal{R}(t)$ we must evaluate the operator exponential in Eq.~\eqref{eq:evolution}. This can be done in the short- and long-time limits, leading to precise asymptotic results for $\mathcal{R}(t)$. As we show below, the asymptotics of $\mathcal{R}(t)$ provide an excellent characterisation of the dynamics, also at finite $t$.

For short times $t\ll1$, the logarithmic ratio \eqref{eq:rlog} reads
\begin{equation}\label{eq:shorttime}
    \mathcal{R}(t) \sim\mathcal{\dot R}(0) t = \frac{\mathcal{\dot F}_h(0)-\mathcal{\dot F}_c(0)}{\mathcal{F}_0}t\,,
\end{equation}
where the dot denotes a time derivative and $\mathcal{F}_0$ is the initial free-energy difference given in Eq.~\eqref{eq:f0}. Through $\mathcal{\dot F}_i(0) = \int_{-\infty}^\infty\!\!\text{d}x\partial_t\,p_i(x,0)\ln[p_i(x,0)/p_f(x)]$, the short-time evaluation of $\mathcal{R}(t)$ depends on the time derivative $\partial_t p(x,0)$, evaluated at $t=0$. By expanding the exponential in Eq.~\eqref{eq:evolution} for $t\ll1$, we obtain $\partial_tp_i(x,0) =(1-\tcal_i)\partial_x^2p_i(x,0)$,
leading us, after integration by parts, to the following integral expression for $\mathcal{\dot F}_i(0)$:
\begin{equation}\label{eq:fdot0}
    \mathcal{\dot F}_i(0) = -\frac{(1-\tcal_i)^2}{\tcal_i}\int_{-\infty}^\infty\text{d}x\, p_i(x,0)V''(x)\,.
\end{equation}
Evaluating Eqs.~\eqref{eq:fdot0} and \eqref{eq:f0} for $i=h,c$, we obtain $\mathcal{R}(t)$ in the short-time limit; see Eq.~\eqref{eq:shorttime}.  For $\sigma=0$ we solve Eq.~\eqref{eq:fdot0} explicitly, which gives
\begin{equation}\label{eq:dtf0sol}
	\mathcal{\dot F}_i(0)=(1-\alpha)\frac{(\tcal_i-1)^2}{\tcal_i^{2/\alpha}}\frac{\Gamma(1-1/\alpha)}{\Gamma(1+1/\alpha)}\,,
\end{equation}
where $\Gamma(x)$ denotes the gamma function \cite{dlmf}. According to Eq.~\eqref{eq:shorttime}, whether the hot or the cold system relaxes faster at short times is determined by the sign of $\mathcal{\dot R}(0)$. As a function of $\alpha$ and $\tcal_h$, where $\tcal_c$ follows from $\mathcal{F}$ equidistance, it is therefore instructive to draw a ``phase diagram,'' marking the different regions in parameter space of initially faster heating [$\mathcal{\dot R}(0)>0$] and initially faster cooling [$\mathcal{\dot R}(0)<0$]. 

\begin{figure}
    \centering
    \includegraphics{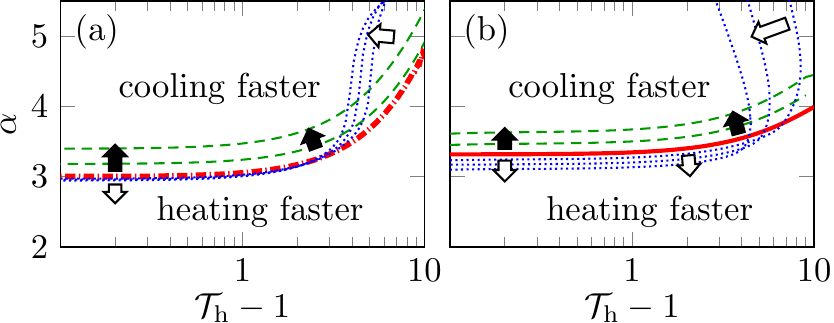}
    \caption{Relaxation-speed phase diagram for short and long times. (a) Short-time phase diagram, calculated from Eq.~\eqref{eq:fdot0}. Critical line for $\sigma=0$ (dash-dotted line), $\sigma=0.5$ and $1$ (dashed lines), and $\sigma=-0.1$, $-0.15$, and $-0.2$ (dotted lines). The black and white arrows indicate how the critical line changes as $\sigma$ is increased and decreased, respectively, from zero. (b) Long-time phase diagram, calculated using the eigenvalue decomposition \eqref{eq:longtime}. Critical line for $\sigma=0$ (solid line), $\sigma=0.2$ and $0.4$ (dashed lines), and $\sigma=-0.1$, $-0.15$, and $-0.2$ (dotted lines). As in Fig.~\ref{fig:phasediag}(a), the black and white arrows indicate how the critical line changes.}
    \label{fig:phasediag}
\end{figure}

Figure~\ref{fig:phasediag}(a) shows the short-time phase diagram for $\sigma=0$, spanned by $\tcal_h$ and $\alpha$. It separates into an upper and a lower part with different short-time behaviours. In the lower part, $\mathcal{\dot R}(0)>0$ so that $\mathcal{R}(t)$ is initially positive for all pairs $\tcal_i$; heating is faster than cooling. In the upper part, cooling is initially faster than heating. The two parts are separated by a critical line (red, dash-dotted line) where $\mathcal{\dot R}(0)=0$ so that $\mathcal{R}(t)$ vanishes to first order in time, $\mathcal{R}(t)\sim \mathcal{O}(t^2)$. For $\sigma=0$, the critical line is obtained by equating $\mathcal{\dot F}_c(0)=\mathcal{\dot F}_h(0)$, and solving for $\alpha$. The smallest critical $\alpha$ value is found to be $\alpha=3$, approached for infinitesimal temperature quenches, $\tcal_i\to1$. We note that this value, and the location of the critical line in general, depends on the choice of temperature measure $\mathcal{F}$. However, the existence of the critical line is robust against moderate changes of $\mathcal{F}$; see Sec.~II of the SM~\cite{SM}.

Similarly, small variations of $\sigma$ away from zero leave the the topology of the short-time phase diagram unchanged. The generic effect of $\sigma>0$ on the critical line is shown by the green, dashed lines and black arrows in Fig.~\ref{fig:phasediag}(a), for values of $\sigma$ up to unity. We observe that slightly increasing $\sigma$ moves the critical line to higher values but does not change the phase diagram qualitatively. 

When $\sigma$ is decreased to negative values, a more complex behaviour emerges, shown by the blue, dotted lines and white arrows in Fig.~\ref{fig:phasediag}(a): For initial temperatures close to equilibrium $\tcal_i\approx1$, the critical line decreases slightly, to $\alpha$ values below $3$. For quenches far from equilibrium, on the other hand, the critical line shifts to higher $\alpha$.

We now turn to the analysis of the long-time limit $t\gg1$ which requires different methods. When the spectrum of $\mathcal{L}$ is discrete, the relaxation of the probability densities $p_i$ to $p_f$ is exponential in the long-time limit. As a result, the densities $p_i$ are determined by the leading right eigenfunctions of $\mathcal{L}$ and their corresponding eigenvalues~\cite{lu2017nonequilibrium}, obtained from the non-Hermitian eigenvalue problem
\begin{equation}\label{eq:eigenvalue}
    \mathcal{L}r_\mu = \lambda_\mu r_\mu\,, \qquad \mathcal{L}^\dagger l_\mu = \lambda_{\mu} l_\mu\,,
\end{equation}
where $l_\mu$ and $r_\mu$ are the left and right eigenfunctions, respectively, and $\lambda_\mu$ with $\lambda_0=0>\lambda_1>\lambda_2>\ldots$ are the associated eigenvalues. Note that the right eigenfunction $r_0$ with eigenvalue $\lambda_0=0$ is given by the steady-state distribution $r_0=p_f$ and $l_0=1$. The eigenfunctions form a complete biorthogonal basis with orthonormality relations
\begin{align}
	\langle l_\mu|r_\nu\rangle = \int_{-\infty}^\infty\!\!\text{d}x\,l_\mu(x) r_\nu(x) = \delta_{\mu\nu}\,.
\end{align}
Expanding $p_i$ in Eq.~\eqref{eq:evolution} in the right eigenbasis of $\mathcal{L}$ we obtain in the long-time limit $t\gg1$,
\begin{equation}\label{eq:longtime}
    p_i(x,t)\sim p_f(x)+ c_{i,\mu} e^{\lambda_\mu t}r_\mu(x)\,,
\end{equation}
where $\mu$ is the lowest number for which $c_{i,\mu}\equiv\langle l_\mu|p_i(0)\rangle \neq 0$. Because our problem is symmetric with respect to the parity operation $x\to-x$, $c_{i,1}$ vanishes, so that $\mu=2$; see Sec.~III in the SM~\cite{SM} for the case of a harmonic potential. All higher-order terms in Eq.~\eqref{eq:longtime} that play a role at finite times are exponentially suppressed in the long-time limit considered here. Using Eqs.~\eqref{eq:rlog} and \eqref{eq:longtime} we find that $\mathcal{R}(t)$ approaches a constant $\mathcal{R}_\infty$ for $t\gg1$ that depends only on the coefficients $c_{i,2}$:
\begin{align}\label{eq:ltime}
	\mathcal{R}(t) \sim 2\ln\left(		\left|\frac{c_{h,2}}{c_{c,2}}	\right|\right)\equiv \mathcal{R}_\infty\,.
\end{align}
Hence, the relative magnitude of the free-energy differences is determined by the initial overlap between the left eigenvector $l_2$ of $\mathcal{L}$ and the initial distributions $p_i(x,0)$ before the temperature quench~\cite{lu2017nonequilibrium}. 

We determine $c_{i,2}$ by solving the eigenvalue problem \eqref{eq:eigenvalue} numerically, discretising it on an evenly spaced, finite lattice with small lattice spacing. Equations \eqref{eq:eigenvalue} then become matrix eigenvalue problems involving large, non-symmetric matrices, whose left and right eigenvectors are approximations of the left and right eigenfunctions $r_\mu$ and $l_\mu$. 

Figure~\ref{fig:phasediag}(b) shows the long-time phase diagram for $\sigma=0$ obtained from numerically computing $c_{i,2}$ and evaluating $\mathcal{R}_\infty$ in Eq.~\eqref{eq:ltime}. The general structure of the long-time phase diagram is qualitatively similar to that of the short-time phase diagram in Fig.~\ref{fig:phasediag}(a), featuring regions of faster heating ($\mathcal{R}_\infty>0$) and faster cooling ($\mathcal{R}_\infty<0$). For long times, however, the critical line [solid line in Fig.~\ref{fig:phasediag}(b)] is located at slightly higher values. Consequently, the minimum of the critical line, attained for close-to-equilibrium quenches, takes the slightly larger value $\alpha\approx3.31$. As in the short-time limit, the long-time critical line is only weakly perturbed by moderate changes of the temperature measure $\mathcal{F}$; see Sec.~II of the SM~\cite{SM} for details.

Upon increasing $\sigma$, we again observe no qualitative change of the phase diagram; the critical line is merely pushed to larger $\alpha$ values [green, dashed lines in Fig.~\ref{fig:phasediag}(b)]. Negative $\sigma$, on the other hand, leads to a qualitative change: For $\sigma<0$, the region of asymptotically faster cooling becomes finite and is completely enclosed by the critical line [blue, dotted lines in Fig.~\ref{fig:phasediag}(b)]. The sensitive dependence of the relaxation dynamics on negative values of $\sigma$, observed both in the short- and long-time limits, must be due to the emergence of bistability of the potential $V(x)$, Eq.~\eqref{eq:pot}. The existence of two potential minima gives rise to multiple relaxation timescales associated with the relaxation within the same minimum and across the two minima.

From the general structure of the phase diagrams we conclude that asymptotically steep potentials (large $\alpha$) lead to faster cooling, compared to $\mathcal{F}$-equidistant heating, when the initial temperature differences are not too large. For small $\alpha$, the opposite is true. Intuitively, this may be explained by noting that for an initially hot system, more probability is located in the tails of the distribution. The steeper the potential, the faster this tail probability is advectively transported toward the potential minimum, leading to faster cooling. For small $\alpha$, this advection effect is weaker, so that it is outperformed by the diffusive broadening of the bulk of the distribution of the cold system, thus resulting in faster heating.

Our analysis reveals the existence of distinct critical lines in the short- and long-time limits. This results in an overlap between the faster-heating and faster-cooling regions at short and long times, giving rise to a crossover region in the phase diagram. In the crossover region, the hot system initially relaxes faster [$\mathcal{R}(t)<0$], but is eventually overtaken by the initially colder system [$\mathcal{R}(t)>0$]. Hence, there must be at least one finite time $t_c>0$ where $\mathcal{R}(t_c)=0$, i.e., the system crosses over from faster cooling to faster heating.

\begin{figure}
    \centering
    \includegraphics{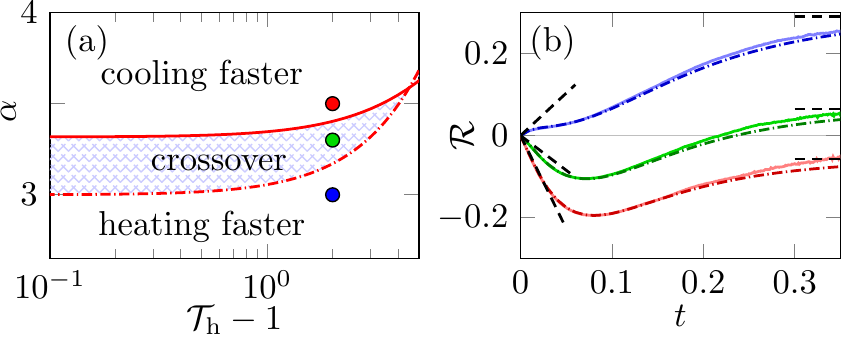}
    \caption{(a) Superimposed short- and long-time phase diagrams for $\sigma=0$, featuring the critical lines in the short-time (dash-dotted line) and long-time (solid line) limits. The crossover region is shown by the cross-hatched region. The coloured dots correspond to the parameter values for the plots in Fig.~\ref{fig:rft}(b).
    (b) $\mathcal{R}(t)$ from different numerical methods for $\tcal_h=3$ and $\alpha=3$, $3.3$ and $3.5$, in blue, green and red, respectively. The dash-dotted lines show results obtained from Eq.~\eqref{eq:evolution}, by numerically calculating the spectrum of $\mathcal{L}$. The solid lines are computed from numerical simulations of the Langevin equation. The black, dashed lines correspond to the short- and long-time asymptotics.}\label{fig:rft}
\end{figure}
Figure~\ref{fig:rft}(a) shows the superimposed short- and long-time phase diagrams for $\sigma=0$ featuring the crossover region (cross-hatched area). The dash-dotted and solid lines show the critical lines from Figs.~\ref{fig:phasediag}(a) and (b), respectively.

In order to study the behaviour of $\mathcal{R}(t)$ in the crossover region, and to validate our previous results, we perform a numerical analysis of the finite-time evolution of $\mathcal{R}(t)$. We focus on a few points in the phase diagram, shown as the differently coloured dots in Fig.~\ref{fig:rft}(a), where we expect qualitatively different behaviours: For the parameter sets represented by the blue and red dots, we expect heating and cooling, respectively, to be faster, both for short and for long times. By contrast, for the parameters of the green dot we expect at least one finite-time crossover from faster cooling to faster heating.

For the finite-time analysis we use two different numerical methods. First, we obtain an approximation of $p_i(x,t)$ by using the discretised analogue of Eq.~\eqref{eq:evolution} obtained with the discretisation scheme discussed earlier. 

The second method approximates the probability density $p_i(x,t)$ by means of a Langevin approach~\cite{van1992stochastic}: We simulate a large number of trajectories $x_i(t)$, $i=h,c$, following the dynamics $\dot x_i(t) = -V'(x) + \xi(t)$, where $\xi(t)$ is a Gaussian white-noise signal with correlation function $\langle \xi(t)\xi(t')\rangle=2\delta(t-t')$. The initial values $x_i(0)$ are sampled from the equilibrium distributions $p_i(x,0)$ prior to the temperature quench. The Langevin equation is solved numerically using an Euler-Maruyama scheme~\cite{kloeden1992stochastic} with a small time step. The probability densities $p_i(x,t)$ are then computed by generating histograms over all locations $x_i(t)$ at discrete times $t$.  

These methods, whose parameters are summarised in Sec.~IV of the SM \cite{SM}, yield two independent numerical approximations $p_i(x,t)$ from which we then calculate $\mathcal{R}(t)$. Figure~\ref{fig:rft}(b) shows the so-obtained $\mathcal{R}(t)$, where the colours of the curves correspond to the colours of the dots in Fig.~\ref{fig:rft}(a). The dash-dotted lines show $\mathcal{R}(t)$ calculated from the discretised operator $\mathcal{L}$. The lighter, solid lines show the corresponding results from the Langevin approach. Also shown are the short- and long-time asymptotes (dashed lines). We observe that the asymptotes represent a good characterisation of the dynamics of $\mathcal{R}(t)$ for all times. In particular, there are no finite-time crossings $\mathcal{R}(t_c)=0$ for the parameter values outside of the crossover region in Fig.~\ref{fig:rft}(a), i.e., for the blue and red curves. Inside the crossover region [see green curve in Fig.~\ref{fig:rft}(b)] we observe only a single crossing.

Furthermore, there is good agreement between the results from the different numerical methods and the asymptotic results. Note that the deviations between the equally coloured curves become larger for longer times. The reason is that for long times, the individual free-energy differences $\mathcal{F}_i(t)$ in Eq.~\eqref{eq:rlog} become exponentially small, so that the relative errors increase as $t$ becomes large. Due to this numerical difficulty, we were unable to evaluate $\mathcal{R}(t)$ until convergence, as can be seen by the discrepancy between our numerical results and the long-time asymptotics [horizontal, dashed lines in Fig.~\ref{fig:rft}(b)].

Finally, we note that far from equilibrium, for $\tcal_c\approx0.0229$ and $\tcal_h\approx5.50$, the short- and long-time critical lines cross [see Fig.~\ref{fig:rft}(a)] which implies the existence of an inverted crossover region very far from equilibrium where heating is initially faster but asymptotically slower than cooling.

In conclusion, $\mathcal{F}$-equidistant thermal relaxation of overdamped diffusions in anharmonic potentials $V(x)$ allows for both faster heating and faster cooling, even when $V(x)$ has a single minimum. As a consequence, the short- and long-time phase diagrams [Figs.~\ref{fig:phasediag}(a) and (b)], spanned by the $(\tcal_h,\alpha)$-parameter space, are nontrivial, exhibiting regions of faster heating and faster cooling. Both for short and for long times, we found that cooling is faster than heating for sufficiently large $\alpha$, and heating is faster than cooling for small $\alpha$. This can be explained in terms of a competition between the advective relaxation of the tail probability of the hot system, and the diffusive broadening of the bulk-probability in the cold system. Despite the similarities between the short- and long-time phase diagrams, we found that their critical lines are different, and that the faster-heating and faster-cooling regions overlap. Superimposing the two, we localised a crossover region [Fig.~\ref{fig:rft}(a)] where cooling is initially faster but the rate of heating eventually overtakes. Outside of the crossover region, we found no crossings, suggesting that the short- and long-time asymptotics faithfully characterise the relative relaxation speeds. The critical lines separating the parameter regions with different behaviours are only weakly perturbed by moderate changes of the temperature measure $\mathcal{F}$ or by an additional harmonic term in the potential $V(x)$, as long as the latter remains single-well, i.e., $\sigma>0$.

It would be interesting to test the relaxation-speed crossover in experiments and thus to reproduce our phase diagram under experimental conditions. This requires tracking the changes in energy and entropy of the system throughout the experiment which is possible in state-of-the-art setups~\cite{kumar2020exponentially,ciliberto2017experiments}. On the theoretical side, it would be desirable to understand the precise dynamical origin of the different relaxation behaviours \footnote{In Sec.~V of~\cite{SM}, we trace back the long-time relaxation asymmetry in harmonic potentials to the convexity of the inverse of the temperature measure $\mathcal{F}$.}. This might lead to optimisation methods for the potential to achieve faster heating or cooling, perhaps in the spirit of first-passage time optimisation \cite{palyulin2012finite,chupeau2020optimizing}.
\begin{acknowledgments}
We thank John Bechhoefer and Massimiliano Esposito for discussions. Funding from the European Research Council within the project ``NanoThermo"  (ERC-2015-CoG  Agreement No.  681456) and the Foundational Questions Institute within the project ``Information as a fuel in colloids and superconducting quantum circuits" (Grant No. FQXi-IAF19-05) is gratefully acknowledged.
\end{acknowledgments}
\end{document}


\title{Supplemental Material for: Relaxation-speed crossover in anharmonic potentials}

\author{Jan Meibohm}
\affiliation{Department of Physics and Materials Science, University of Luxembourg, L-1511 Luxembourg, Luxembourg}
\author{Danilo Forastiere}
\affiliation{Department of Physics and Materials Science, University of Luxembourg, L-1511 Luxembourg, Luxembourg}
\author{Tunrayo Adeleke-Larodo}
\affiliation{Department of Physics and Materials Science, University of Luxembourg, L-1511 Luxembourg, Luxembourg}
\author{Karel Proesmans}
\affiliation{Department of Physics and Materials Science, University of Luxembourg, L-1511 Luxembourg, Luxembourg}
\affiliation{Hasselt University, B-3590 Diepenbeek, Belgium}

\begin{abstract}
In this Supplemental Material, we discuss how the relaxation asymmetry after a $\mathcal{F}$-equidistant temperature quench relates to the Markovian Mpemba effect described in the literature. We also we show that the short- and long-time phase diagrams are robust against changes in the temperature measure, by considering a generalisation $\mathcal{F}^q$ of the free-energy difference $\mathcal{F}$ discussed in the main text. Further, we outline a computation of the $c_2$-coefficient for over- and underdamped Langevin systems in harmonic potentials, and explain our numerical method.  Finally, we show how in the long-time limit and for harmonic potentials, the heating-cooling asymmetry can be traced back to a convexity property of the inverse of the temperature measure.
\end{abstract}
\maketitle
\section{Connection with Markovian Mpemba effect}
In this section, we discuss the relation between the relaxation asymmetry in $\mathcal{F}$-equidistant quenches and the Markovian Mpemba effect. Both effects refer to the thermal relaxation of two identical systems at different initial temperatures towards the same final temperature which allows us to use similar methods for their description: In the main text, we used the method of calculating the coefficients $c_{2}$ to determine the long-time behaviour of the cold and hot systems after an $\mathcal{F}$-equidistant temperature quench. This approach for obtaining the asymptotic rates of heating and cooling after temperature quenches has been previously applied to the Markovian Mpemba effect in Ref.~\cite{lu2017nonequilibrium}. 

The perhaps most obvious difference between the two effects is the ordering of the temperature ratios. For the $\mathcal{F}$-equidistant quench we have $\tcal_c<1<\tcal_h$. For the Markovian Mpemba effect, on the other hand, one has $1<\tcal_c<\tcal_h$, or $\tcal_c<\tcal_h<1$ for the inverse Mpemba effect \cite{lu2017nonequilibrium}. It was shown in Ref.~\cite{lu2017nonequilibrium} that a sufficient condition to observe the Mpemba effect is that $c_2(\tcal)$ is a non-monotonic function of the initial temperature ratio $\tcal>1$. In this case, there always exist temperature pairs $1<\tcal_c<\tcal_h$ so that $|c_2(\tcal_c)|>|c_2(\tcal_h)|$. As a result, the initially hotter system (with initial temperature $\tcal_h$) relaxes asymptotically faster than the colder one (with initial temperature $\tcal_c$).

The condition $\mathcal{F}_c(t)<\mathcal{F}_h(t)$ for all times, as discussed in the main text and in Refs.~\cite{lapolla2020faster,manikandan2021faster}, is somewhat stronger, as it is supposed to hold also at finite times. For the sake of comparison with the Mpemba effect, we here restrict ourselves to the long-time limit, where the $c_2$-coefficient is a faithful measure for both phenomena.
\begin{figure}
    \centering
    \includegraphics{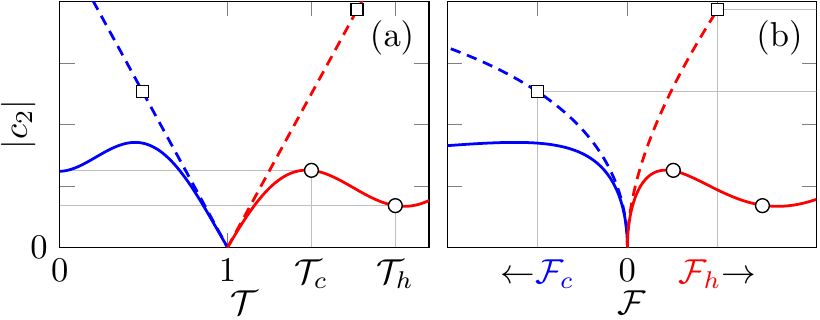}
    \caption{The Markovian Mpemba effect and the long-time limit of the $\mathcal{F}$-equidistant temperature quench. (a) Schematic of the magnitude of $c_2$ as a function of the initial temperature $\tcal$. The blue and red solid lines show a hypothetical $c_2(\tcal)$. The circles correspond to a temperature pair that exhibits the Markovian Mpemba effect, the squares show an $\mathcal{F}$-equidistant temperature pair [see Fig.~\ref{fig:mpemba}(b)]. The dashed lines showcase a linear $c_2(\tcal)$. (b) Same schematic as in Fig.~\ref{fig:mpemba}(a), but as a function of the free-energy difference $\mathcal{F}$. The squares represent a pair of $\mathcal{F}$-equidistant temperatures.}
    \label{fig:mpemba}
\end{figure}

Figure~\ref{fig:mpemba}(a) schematically shows $|c_2|$ as a function of $\tcal$. Here, the Markovian Mpemba effect corresponds to the non-monotonicity of the red branch: We may find pairs of initial temperatures $\tcal_c$ and $\tcal_h$ (white circles) so that $|c_2(\tcal_c)|>|c_2(\tcal_h)|$. The equivalent phenomenon, but for $\mathcal{T}<1$ (blue solid line), corresponds to the inverse Markovian Mpemba effect.

Figure~\ref{fig:mpemba}(b) schematically shows $|c_2|$ as a function of $\mathcal{F}$ which characterises $\mathcal{F}$ equidistance temperature quenches in the long-time limit. The initial temperature ratios $\tcal_c$ and $\tcal_h$ are chosen $\mathcal{F}$-equidistant such that $\mathcal{F}(\tcal_c)=\mathcal{F}(\tcal_h)$, see gray lines in Fig.~\ref{fig:mpemba}(b). Loosely speaking, the initial temperatures for Markovian Mpemba effect always lie on either branch (the red or the blue) in Fig.~\ref{fig:mpemba}, where the red corresponds to the Mpembda effect and the blue to its inverse. For the $\mathcal{F}$-equidistant quench, initial temperatures from two different branches, red and blue, are chosen.

Another important difference is that the Markovian Mpemba effect relies only on the order of the temperatures, and on how this ordering changes under the action of the mapping $|c_2|:\tcal\mapsto |c_2(\tcal)|$. In particular, this means that the occurrence of the Mpemba effect is invariant under any mapping of the initial temperatures that preserves their ordering \cite{lu2017nonequilibrium}.

In our example in Fig.~\ref{fig:mpemba}, this can be readily verified. The free-energy difference $\mathcal{F}$ preserves the temperature ordering, i.e., $1<\tcal_c<\tcal_h$ implies $0<\mathcal{F}(\tcal_c)<\mathcal{F}(\tcal_h)$, so that also $|c_2[\mathcal{F}(\tcal_c)]|>|c_2[\mathcal{F}(\tcal_h)]|$ [see circles in Figs.~\ref{fig:mpemba}(a) and (b)]. Hence, the Mpemba effect is preserved by this mapping. The same logic applies for the inverse Mpemba effect.

On the other hand, $\mathcal{F}$ equidistance does depend on the choice of the specific temperature measure $\mathcal{F}$, because this notion relies on the actual values of the distances and not just on their ordering. The dashed lines in Figs.~\ref{fig:mpemba}(a) and (b) show how the notion of $\mathcal{F}$ equidistance is transformed by the mapping $\mathcal{F}\mapsto\tcal(\mathcal{F})$ for a $c_2$ that is linear in $\tcal$,
\begin{equation}\label{eq:c2lin}
    |c_2(\tcal)|\propto |\tcal-1|\,,
\end{equation}
see Sec.~\ref{sec:c2_quadratic} for an example. With $c_2$ as in Eq.~\eqref{eq:c2lin}, a $\tcal$-equidistant temperature pair, i.e., with $|\tcal_c-1|=|\tcal_h-1|$, $\tcal_c<1<\tcal_h$, has $|c_2(\tcal_c)| = |c_2(\tcal_h)|$. However, if we choose $\mathcal{F}$-equidistant temperatures instead [see squares in Fig.~\ref{fig:mpemba}(b)], we generically have $c_2(\tcal_c)\neq c_2(\tcal_h)$ because then $|\tcal_c-1|\neq|\tcal_h-1|$ [see squares in Fig.~\ref{fig:mpemba}(a)]. In other words, $\mathcal{F}$ equidistance does not imply $\tcal$-equidistance, nor equidistance with respect to any other generic temperature measure. Note that this argument also holds for a generic, nonlinear $c_2(\tcal)$.

In summary, although the setup for the two effects is similar, there are two important differences: The temperature ordering and the dependence on the temperature measure.
\section{Robustness of the phase diagram}
In this section, we show that the short- and long-time phase diagrams [Fig.~2(a) and 2(b) in the main text] are robust against moderate changes of the temperature measure. To this end, we study the effect of replacing the free-energy difference $\mathcal{F}$ by the ``deformed free-energy difference" $\mathcal{F}^q$,
\begin{align}\label{eq:fqdef}
	\mathcal{F}_i^q(t) = \frac1{q(q-1)}\int_\infty^\infty\!\!\!\text{d}x\left\{\left[\frac{p_i(x,t)}{p_f(x)}\right]^{q-1}-1\right\} p_i(x,t)\,,
\end{align}
which reduces to $\mathcal{F}$, as defined in the main text, in the limit $q\to1$. This can be seen by noting that for $|q-1|\ll1$,
\begin{align}
    \left[\frac{p_i(x,t)}{p_f(x)}\right]^{q-1}\!\!-1\sim(q-1)\ln  \left[\frac{p_i(x,t)}{p_f(x)}\right]\,.
\end{align}
The limit $q\to0$, on the other hand, corresponds to $\mathcal{F}$ when swapping $p_i(x,t)$ and $p_f(x)$ which follows from
\begin{align}
    \left[\frac{p_i(x,t)}{p_f(x)}\right]^{q-1}\sim\frac{p_f(x)}{p_i(x,t)}\left[1+q\ln\frac{p_i(x,t)}{p_f(x)}\right]\,,
\end{align}
for $|q|\ll1$. The deformed free-energy difference $\mathcal{F}^q$ has a probabilistic interpretation in terms of a generalisation of the Kullback-Leibler divergence to Tsallis statistics \cite{furuichi2004fundamental,taneja2004generalized,huang2016generalization}. For our purpose, $\mathcal{F}^q$ is convenient because it  encompasses a rather broad range of temperature measures while allowing for similar analytical progress as the original free-energy difference $\mathcal{F}$, attained for $q=1$. For simplicity, we consider only the case $\sigma=0$ in what follows.

In order to relate the initial and final temperature ratios, we first need to compute $\mathcal{F}^q_0\equiv \mathcal{F}^q(0)$ as function of $\tcal$. Performing the integral in Eq.~\eqref{eq:fqdef} for the initial, stationary distributions we find
\begin{align}\label{eq:fq0}
	\mathcal{F}^q_0 = \frac1{q(q-1)}\left[\left(\frac{\tcal_q}{\tcal^q}\right)^{1/\alpha} -1			\right]\,,
\end{align}
where we defined
\begin{align}\label{eq:tq}
	 \tcal_q \equiv \frac{\tcal}{1-(q-1)(\tcal-1)}\,.
\end{align}
For $q\to1$, we have $\mathcal{F}^q_0\to\mathcal{F}_0$ given in Eq.~(9) in the main text. In order for $\mathcal{F}^q_0$ to remain finite and real, we must require $\tcal_q>0$ which in turn requires $\tcal_h<q/(q-1)$ for $q>1$. The limit $\tcal_h\to q/(q-1)$ corresponds to $\tcal_c\to0$ for the cold temperature ratio.
\begin{figure*}
	\centering
	\includegraphics{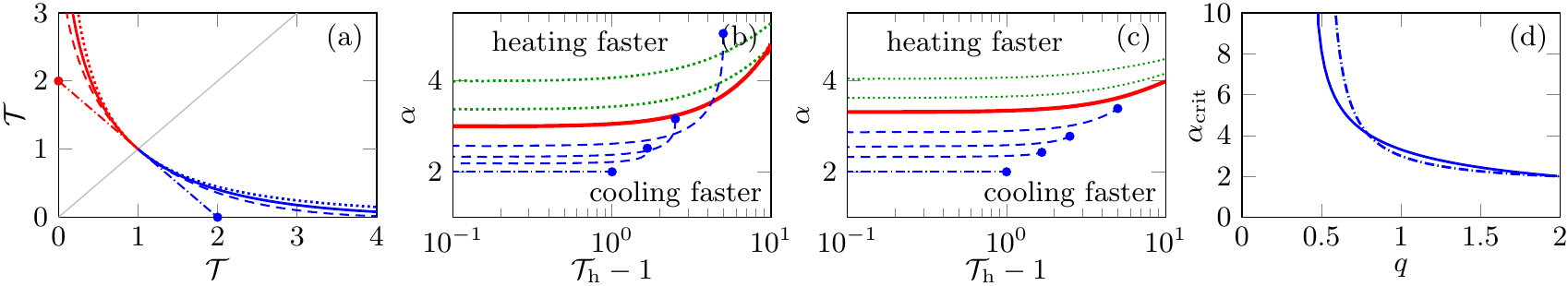}
\caption{(a) Temperatures $\tcal_c(\tcal_h)$ (red) and $\tcal_h(\tcal_c)$ (blue) for $q=1$ (solid), $q=0.8$ (dotted), $q=1.2$ (dashed), and the linear measure attained for $q=2$ (dash-dotted). The dots show the cases when $\tcal_h\to q/(q-1)$ and $\tcal_c\to0$. (b) Short-time phase diagram for different values of $q$. The critical line for $q=1$, the case discussed in the main text,  is shown as the red, solid line. The green, dotted lines correspond to $q<1$, with values $q=0.9$ and $q=0.8$. The blue, dashed lines correspond to $q>1$ with values $q=1.2$, $1.4$ and $1.6$. The dash-dotted line shows the critical line for the linear measure attained for $q=2$. The blue dots show the critical points for the limits $\tcal_h\to q/(q-1)$ and $\tcal_c\to0$. (c) Long-time phase diagram for different values of $q$. Same $q$ values and line styles as in (b). (d) Critical $\alpha$ for close-to-equilibrium quenches in the short- (dash-dotted) and long-time (solid) limits as functions of $q$.}\label{fig:tmeasures}
\end{figure*}

We now use Eq.~\eqref{eq:fq0} to relate the hot and cold temperatures, thus giving $\tcal_h(\tcal_c)$ and $\tcal_c(\tcal_h)$. Figure~\ref{fig:tmeasures}(a) shows $\tcal_h(\tcal_c)$ and $\tcal_c(\tcal_h)$, obtained from combining both temperature branches as in Fig.~1(b) in the main text, for different $q$. Also shown is the case of a linear temperature measure (dash-dotted line). We observe that varying $q$ leads to a different relations between the hot and cold temperatures, as expected. Larger $q>1$ leads to a $g$ that is closer to that of the linear temperature measure $|\tcal_h-1|=|\tcal_c-1|$ [dotted line in Fig.~\ref{fig:tmeasures}(a)]. Interestingly, a straightforward calculation using Eq.~\eqref{eq:fq0} shows that for $q=2$, $\mathcal{F}^q$-equidistance becomes equivalent to $\tcal$-equidistance (dash-dotted line), i.e., to $|\tcal_h-1|=|\tcal_c-1|$. Conversely for $q<1$, the temperature measure $\mathcal{F}^q$ is further away from the linear measure.
\subsection{Short-time phase diagram}
The short-time phase diagram for the deformed free-energy difference \eqref{eq:fq0} is obtained by computing $\mathcal{R}^q(t)\equiv\ln[\mathcal{F}^q_h(t)/\mathcal{F}_c^q(t)]$ for short times using the equivalent of Eq.~(12) in the main text. This requires evaluating $\mathcal{\dot F}^q(0)$, given by
\begin{align}\label{eq:dfq0}
	\mathcal{\dot F}^q(0) = -(\alpha-1)\frac{(1-\tcal)^2}{(\tcal_q \tcal^q)^{1/\alpha}}\left(\frac{\tcal_q}{\tcal}\right)^2\frac{\Gamma(1-1/\alpha)}{\Gamma(1+1/\alpha)}\,,
\end{align}
which for $q\to1$ reduces to Eq.~(14) in the main text. Equation~\eqref{eq:dfq0} allows us to evaluate $\mathcal{\dot R}^q(0) = [\mathcal{\dot F}_h^q(0)-\mathcal{\dot F}_c^q(0)]/\mathcal{F}^q_0$, from which we obtain the short-time phase diagram, marking regions in parameter space where $\mathcal{\dot R}^q(0)>0$ (faster heating) and $\mathcal{\dot R}^q(0)<0$ (faster cooling). Figure~\ref{fig:tmeasures}(b) shows the critical lines of the phase diagram for different $q$. We observe that $q<1$ generically pushes the critical line to higher $\alpha$ values, while $q>1$ decreases it. However, the topology of the phase diagram remains intact in a whole neighbourhood of $q$ values around $q=1$. We therefore conclude that the short-time phase diagram is robust against moderate changes of the temperature measure, at least against those of the kind given by Eq.~\eqref{eq:fqdef}.
\subsection{Long-time phase diagram}
In the long-time limit, $\mathcal{R}^q(t)$ simplifies to
\begin{align}\label{eq:rqlong}
	\lim_{t\to\infty}\mathcal{R}^q(t) = 2\ln\left(\left|\frac{c_{2,h}}{c_{2,c}}\right|\right)=\mathcal{R}_\infty\,,
\end{align}
independent of $q$ and thus identical to the expression in Eq.~(18) in the main text. This is not unexpected, as it can be shown that a broad class of temperature measures is equivalent in the long-time limit~\cite{lu2017nonequilibrium}. This does, however, not mean that the long-time phase diagram is unaffected by the change $\mathcal{F}\to\mathcal{F}^q$. The reason is that the initial temperature ratios, $\tcal_h$ and $\tcal_c$ are still related to each other by $\mathcal{F}^q$-equidistance, which is in general different from $\mathcal{F}$ equidistance [see Fig.~\ref{fig:tmeasures}(a)].

As in the case $q=1$ (see main text), we compute $\mathcal{R}_\infty$ by calculating $c_2(\tcal)$ numerically from the spectral decomposition of the Fokker-Planck operator. From this, we obtain the long-time phase diagram by marking regions in the $(\tcal_h,\alpha)$ plane where $\mathcal{R}_\infty>0$ (heating is faster than cooling) and $\mathcal{R}_\infty<0$ (cooling is faster than heating). Figure~\ref{fig:tmeasures}(c) shows the long-time phase diagram for different values of $q$. As for short times, the long-time phase diagram is robust against moderate changes of $q$ away from $q=1$. We observe that a decrease in $q$ moves the critical $\alpha$ line to larger values. Conversely, increasing $q$ shifts the critical line to lower $\alpha$ values. The general structure of the long-time phase diagram remains unaltered, just as for short times. Interestingly, however, we find that for $q$ values below $q\approx0.78$, the crossover region observed for $q=1$ disappears, giving rise to an extended inverted crossover region (see main text).
\subsection{Critical line for close-to-equilibrium quenches}
We now show how to compute the critical line for close-to-equilibrium quenches, where $|\tcal_i-1|\ll 1$, in both the short- and the long-time limit. To this end, we first calculate $\tcal_c(\tcal_h)$ asymptotically for $\tcal_h-1\equiv\varepsilon\ll1$ by expanding $\mathcal{F}^q_0$ in Eq.~\eqref{eq:fq0} in $\tcal-1$:
\begin{align}\label{eq:fq0asym}
	\mathcal{F}^q_0 \sim \frac{1}{2\alpha}(\tcal-1)^2 + \frac{q-2}{3\alpha}(\tcal-1)^3\,.
\end{align}
Now we make the ansatz $\tcal_c-1 \sim -\varepsilon + A\varepsilon^2$ where $\varepsilon=\tcal_h-1$. We substitute this ansatz into Eq.~\eqref{eq:fq0asym}, expand in small $\varepsilon$ and solve for $A$. This gives the following relation $\tcal_c(\tcal_h)$ valid for small $\tcal_h-1\ll1$:
\begin{align}\label{eq:tcasym}
	\tcal_c -1 	\sim -(\tcal_h-1) + \frac23(2-q)(\tcal_h-1)^2\,.
\end{align}
In the short-time limit, the critical line is determined by the condition $\mathcal{\dot R}(0)=0$. To obtain the critical line for close-to-equilibrium quenches, we expand $\mathcal{\dot R}(0)$ in $\tcal_h-1\ll1$ using Eq.~\eqref{eq:tcasym}. This gives
\begin{align}
	\mathcal{\dot R}(0) \sim -\frac{8}{3}(\alpha -1)[\alpha(2q-1)-3 q]\frac{\Gamma (1-1/\alpha)}{\Gamma (1+1/\alpha)}(\tcal_h-1)\,.
\end{align}
Setting this expression to zero gives the asymptotic value of the critical line for close-to-equilibrium quenches. We obtain
\begin{align}\label{eq:alphashort}
	\alpha \sim \frac{3q}{2q-1}\,.
\end{align}
The dash-dotted line in Fig.~\ref{fig:tmeasures}(d) shows Eq.~\eqref{eq:alphashort} as a function of $q$. For $q=1$ we obtain the value $\alpha=3$ stated in the main text. The value $q=2$, which corresponds to the linear temperature measure, gives critical $\alpha=2$, the harmonic potential. This is the case also in the long-time limit as shown in Sec.~\ref{sec:c2_quadratic}, see discussion below. We also find that the short-time critical line is pushed to infinity as $q\to1/2$, and disappears for $q<1/2$. This shows that the general structure of the short-time phase diagram may change for temperature measures that are very different from $\mathcal{F}$.

In the long-time limit, the critical line is determined by $\mathcal{R}_\infty=0$ and thus by $|c_2(\tcal_h)| = |c_2(\tcal_c)|$. To obtain the critical $\alpha$ value for close-to-equilibrium quenches, we expand $c_2(\tcal)$ around $\tcal-1\ll1$:
\begin{align}
	c_2(\tcal) \sim c'_2(1)(\tcal-1) + \frac{c''_2(1)}2(\tcal-1)^2\,.
\end{align}
Now, using Eq.~\eqref{eq:tcasym} together with $|c_2(\tcal_h)| = |c_2(\tcal_c)|$ we obtain the relation
\begin{align}\label{eq:alphalong}
	c_2''(1) = \frac{2}3(q-2)c_2'(1)\,.
\end{align}
This expression determines the critical $\alpha$ value for close-to-equilibrium quenches in the long-time limit. To evaluate $\alpha$, we numerically compute $c_2'(1)$ and $c_2''(1)$ from $c_2(\tcal)$, along a fine grid of $\alpha$ values. The result is shown as the solid line in Fig.~\ref{fig:tmeasures}(d). We observe that the critical lines in the short- and long-time limit behaves similarly as functions of $q$. For $q=1$ we obtain the long-time value $\alpha\approx3.31$ stated in the main text. For $q=2$, we find $\alpha=2$. This can be understood by noting that $q=2$ in Eq.~\eqref{eq:alphalong} gives $c_2''(1)=0$, which is precisely the case for harmonic potentials, i.e., $\alpha=2$, see Eqs.~\eqref{eq:overdamped_c2} and \eqref{eq:underdamped_c2} in Sec.~\ref{sec:c2_quadratic}. Note that at $q\approx0.78$ the close-to-equilibrium values for $\alpha$ in short- and long-time limits cross, so that the crossover region of the phase diagram vanishes, as discussed before. Furthermore, the long-time critical line is pushed to infinity for $q\approx0.45$, similarly to what we observed in the short-time limit.

To conclude this section, we analysed the change of the phase diagram under variation of the temperature measure by setting $\mathcal{F}\to \mathcal{F}^q$. We observed that the topology of the phase diagram is robust under moderate changes of this measure. Decreasing $q$ too far away from $q=1$ first leads to a loss of the crossover region at $q\approx0.78$, giving rise to an extended inverted crossover region (see main text), followed by the complete vanishing of the critical lines in the short-time (at $q=1/2$) and in the long-time (at $q\approx 0.45$) limit.
\section{Analytical expression of $c_2$ for harmonic potentials}\label{sec:c2_quadratic} 
In this section, we outline the derivation of the $c_2$-coefficient for over- and underdamped diffusion in harmonic potentials. In particular, we show that the $c_2$-coefficients are linear in the temperature ratio $\tcal$. To calculate $c_2$, we need to find the left eigenvectors associated with the Fokker-Planck operator $\mathcal{L}$ [Eq.~(5) in the main text]. 

We start by discussing the overdamped limit. The potential $V(x)$ [Eq.~(4) in the main text] is chosen to be harmonic $V(x)=x^2$, i.e., $\sigma=0$ and $\alpha=2$. For harmonic potentials, the Fokker-Planck equation can be solved analytically~\cite{risken1996fokker}. The first three left eigenfunctions read
\begin{equation}
    l_0(x)=1\,,\qquad l_1(x)=x\,,\qquad l_2(x)=2x^2-1\,.
\end{equation}
The associated eigenvalues are $\lambda_0=0,$ $\lambda_1=-2$ and $\lambda_2=-4$, respectively. With the initial distributions given by $p(\tcal) = \exp(-x^2/\tcal)/\sqrt{\tcal\pi}$, we obtain from
\begin{align}
    c_\mu(\tcal) = \langle l_\mu|p(\tcal)\rangle \label{eq:scalarprod}
\end{align} 
the coefficients 
\begin{equation}\label{eq:overdamped_c2}
    c_0(\tcal)=1\,,\qquad c_1(\tcal) = 0\,,\qquad c_2(\tcal)=\tcal-1\,.
\end{equation}
We observe that $c_2$ is linear in the dimensionless temperature ratio $\tcal=T_{c/h}/T_f$ for overdamped diffusion in a harmonic potential.

The calculation for the underdamped case goes along similar lines. For the computation of the left eigenfunctions of the Kramers equation we refer to, e.g., Appendix A of Ref.~\cite{titulaer1978systematic}. 
From Eq.~\eqref{eq:scalarprod} one finds
\begin{equation}\label{eq:underdamped_c2}
    c_0(\tcal)=1\,,\quad c_1(\tcal) = 0\,,\quad c_{2}(\tcal)=N_2\left(\tcal-1\right)\,, 
\end{equation}
with
\begin{equation}\label{eq:underdamped_N2}
    N_2= 1 + 4\left(\zeta-\sqrt{\zeta^2-1}\right)^2\,.
\end{equation}
Here, $\zeta = \sqrt{m\gamma^2/(8k)}$ is the dimensionless damping ratio with particle mass $m$, friction coefficient $\gamma$ and harmonic coupling $k$. Equations~\eqref{eq:underdamped_c2} show that also in the underdamped limit, the coefficient $c_2$ is linear in the temperature ratio $\tcal$. Upon taking the overdamped limit $\zeta\to\infty$, the second term in Eq.~\eqref{eq:underdamped_N2} vanishes and we recover Eqs.~\eqref{eq:overdamped_c2}.
\section{Numerical method}
\begin{table}
\caption{Parameter values in our numerical simulations.}\label{tab:numerics}
\begin{tabular}{c|c|c|c|c|c|c|c}
	$L$	&	$dx$				&	$N_\text{grid}$		&	$dt$		&	$N_\text{traj}$		&	$N_\text{bin}$	&	$m$		&	$M$\\
	\hline
	10	&	$1.25\times10^{-4}$	&	$8\times10^{3}$	&	$10^{-3}$	&	$1.62\times10^{8}$	&	$10^{4}$		&	$10^{-4}$	&	$10^{2}$
\end{tabular}
\end{table}

Here we explain in more detail the numerical methods used to obtain the results presented in Fig.~3(b) of the main text. The numerical values of all parameters are summarised in Tab.~\ref{tab:numerics}.

For the first method, we discretise the Fokker-Planck operator $\mathcal{L}$ [Eq.~(5) in the main text] on a finite domain $[-L/2,L/2]$ with absorbing boundary conditions, using an evenly spaced grid with $N_\text{grid}$ points, resulting in a grid spacing of $dx = L/(N_\text{grid}-1)$. To obtain the discretised probability density $p_i(x_k,t)$ at grid point $x_k$, $k\in\{1,\ldots,N_\text{grid}\}$, and time $t$, we then numerically compute the matrix exponential in Eq.~(11) of the main text.

For the second method we simulate $N_\text{traj}$ trajectories $\{x_i(t)\}$ using an Euler-Maruyama scheme with time step $dt$. The initial points of each trajectory are sampled from the initial probability densities $p_i(x,0)$ at temperature ratios $\tcal_i$, $i=h,c$. At each time step $t_n$, the probability densities are calculated by binning the positions $x_i(t_n)$ of the trajectories into $N_\text{bin}$ bins with logarithmically spaced edges at locations ranging between $m$ and $M$.
\section{Heating-cooling asymmetry for harmonic potential}\label{sec:measures}
In this section, we show that in the long-time limit and for harmonic potentials, the asymmetry between heating and cooling in $\mathcal{F}$-equidistant temperature quenches can be traced back to the convexity of the inverse of $\mathcal{F}$.

To this end, we consider a generic measure $f(\tcal)$ for the distance from equilibrium, assuming that $f$ has a single minimum at $\tcal=1$ where $f(1)=0$. The function $f$ has two branches, corresponding to $\tcal\leq1$ and $\tcal\geq1$, that we call $f_1$ and $f_2$, respectively. Along the same lines as in Ref.~\cite{lapolla2020faster} and in the main text, we define an $f$-equidistant temperature pair $(\tcal_c,\tcal_h)$ by
\begin{align}\label{eq:fequidistance}
	f_1(\tcal_c) = f_2(\tcal_h)\,.
\end{align}
By inverting $f_1$ and $f_2$, we explicitly relate $\tcal_c$ and $\tcal_h$:
\begin{align}\label{eq:trelation}
	\tcal_c = f_1^{-1}[ f_2(\tcal_h)]\,,\qquad 	\tcal_h = f_2^{-1}[f_1(\tcal_c)]\,.
\end{align}
In order to obtain a single function $g$ that is defined on the positive real line, we join the two branches given in Eq.~\eqref{eq:trelation} according to
\begin{align}\label{eq:geqn}
	g(\tcal) = \begin{cases} f_2^{-1}[f_1(\tcal)]\,,	&	\tcal<1\,,\\	f_1^{-1}[f_2(\tcal)]\,,	&	\tcal\geq1\,.	\end{cases}
\end{align}
It follows from the properties of $f_1$ and $f_2$ that $g(1)=1$, and that $g$ is a decreasing function of $\tcal$. Furthermore, since $(f_2^{-1}\circ f_1)^{-1}=f_1^{-1}\circ f_2$, $g$ is its own inverse, $g(g(\tcal))=\tcal$. Geometrically, the $\tcal>1$-branch of $g$ is the mirror image of the $\tcal<1$-branch, with respect to an axis that goes through the origin at angle $\pi/4$. 

In addition to these generic properties, we make the assumption that $g$ is convex, i.e., $g[t x + (1-t)y]\leq tg(x) + (1-t)g(y)$, for $0\leq t\leq 1$.

Under these assumptions, we now show that for $f$-equidistant temperatures and a linear $c_2(\tcal)$ as in Eqs.~\eqref{eq:overdamped_c2} and \eqref{eq:underdamped_c2}, heating occurs faster than cooling as $t\to\infty$. This corresponds to proving the condition $|\tcal_c-1|< |\tcal_h-1|$, which, using $\tcal_c<1<\tcal_h$ and $\tcal_{c,h} = g(\tcal_{h,c})$, is expressed as
\begin{align}\label{eq:assumption}
	g(\tcal)\geq2-\tcal \,,
\end{align}
where $\tcal=\tcal_c$ for $\tcal<1$ and $\tcal=\tcal_h$ for $\tcal\geq1$. 

We now prove \eqref{eq:assumption} by contradiction. First, we assume that there exists a $\tcal_1<1$ so that $g(\tcal_1)<2-\tcal_1$. The function $g$ maps $\tcal_1$ to a second temperature that we call $\tcal_2=g(\tcal_1)>1$. Due to the convexity of $g$, we have
\begin{align}\label{eq:convex}
	g\left(\frac{\tcal_1+\tcal_2}{2}\right)\leq \frac12 g(\tcal_1) + \frac12 g(\tcal_2)\,.
\end{align}
On the right-hand side of Eq.~\eqref{eq:convex}, $g(\tcal_1)=\tcal_2$ and, since $g$ is its own inverse, $g(\tcal_2) = \tcal_1$. We thus write
\begin{align}
	\frac12 g(\tcal_1) + \frac12 g(\tcal_2) = \frac{\tcal_1+\tcal_2}{2}\,.
\end{align}
Now, by assumption $g(\tcal_1)=\tcal_2<2-\tcal_1$, so that $(\tcal_1+\tcal_2)/2<1$. And, since $g$ is decreasing, $g[(\tcal_1+\tcal_2)/2]>g(1)=1$. Using Eq.~\eqref{eq:convex} we therefore obtain the contradiction
\begin{align}
	1<g\left(\frac{\tcal_1+\tcal_2}{2}\right)\leq \frac{\tcal_1+\tcal_2}{2}<1\,,
\end{align}
which proves that we must have $g(\tcal)\geq 2-\tcal$ for $\tcal<1$. The same reasoning works for $\tcal>1$. For $\tcal=1$ we obtain equality in Eq.~\eqref{eq:assumption}. Putting the different cases together proves Eq.~\eqref{eq:assumption}. 

In summary, under the assumption that $g(\tcal)$ is convex, heating occurs faster than cooling whenever $c_2$ is a linear function of $\tcal$. As shown in Sec.~\ref{sec:c2_quadratic}, the coefficient $c_2$ is linear for over- and underdamped diffusions in harmonic potentials. Furthermore, the function $g$ [Eq.~\eqref{eq:geqn}] that corresponds to the free-energy difference $\mathcal{F}$ is convex [see Fig.~1(b) in main text] in this case. The same holds for deformed free-energy difference $\mathcal{F}^q$, see Fig.~\ref{fig:tmeasures}(a). Hence, for harmonic systems in the long-time limit, the relaxation asymmetry is a consequence of the convexity of $g$, which relates the hot and cold temperatures through the equidistance condition \eqref{eq:fequidistance}.